\documentclass[english]{article}
\usepackage[latin9]{inputenc}
\usepackage{geometry}
\usepackage{amsmath}
\usepackage{setspace}
\makeatletter
\newcommand{\lyxaddress}[1]{
	\par {\raggedright #1
	\vspace{1.4em}
	\noindent\par}
}

\makeatother

\usepackage{babel}
\begin{document}
\title{W-exchange contribution in hadronic decays of bottom baryon}
\author{Fayyazuddin}
\maketitle

\lyxaddress{Department of Physics, Quaid-i-Azam University, Islamabad, Pakistan.}

The nonleptonic decays of $\varLambda_{b}$ that are dominated by
$W$ exchange are studied. In particular, the decay modes $\varLambda_{b}\to\varDelta^{0}D^{0},\varDelta^{-}D^{+},\Sigma^{*-}D_{s}^{+}$,
$\Lambda_{b}\to\Sigma_{c}^{*+}\pi^{-},\Sigma_{c}^{*0}\pi^{0},\Xi_{c}^{*0}K^{0}$
and $\varLambda_{b}\to\Sigma_{c}^{+}\pi^{-}$ are analyzed. In an
another aspect, the decay $\Lambda_{b}\to\Lambda_{c}^{+}\pi^{-}$
in the factorization anstaz is studied. It is shown that factorization
contributes to parity violating (s-wave) amplitude $A$ only. Hence
factorization gives asymmetry parameter $\alpha=0$. However, the
dominant contribution to parity conserving (p-wave) amplitude $B$
comes from $W$ exchange, i.e., from baryon pole, giving asymmetry
parameter $\alpha=-0.77$.

\section{Introduction}
In the standard model (SM), two body hadronic decays for the heavy flavor are analyzed in terms of the effective Hamiltonian \cite{FD-Book, Neubert}
\begin{eqnarray}
\mathcal{H}_{eff} & = & V_{cb}V^{*}_{cs}[a_1(\bar{s}c)_{V-A}(\bar{c}b)_{V-A} + a_2(\bar{c}c)_{V-A}(\bar{s}b)_{V-A}]\label{1na}\\
& = &V_{cb}V^{*}_{ud}[a_1(\bar{d}u)_{V-A}(\bar{c}b)_{V-A} + a_2(\bar{c}u)_{V-A}(\bar{d}b)_{V-A}]\label{1nb}
\end{eqnarray}
where
\begin{eqnarray}
a_1 & = & C_1+\frac{1}{3}C_2  :\quad\quad\quad\quad \text{tree diagram}\notag\\
a_2 & = & C_2+\frac{1}{3}C_1 :\quad\quad\quad\quad \text{color suppressed tree diagram}
\end{eqnarray}
The Hamiltonian \ref{1na} arises from the transition $b\to c+s+\bar{c}$ and \ref{1nb} from the transition $b\to c + d + \bar{u}$. Only the Hamiltonian \ref{1na} is relevant for the decays of bottom baryon $\Lambda_b$ which is studied in this paper. Both these Hamiltonian correspond to the decays which are not Cabibbo suppressed. 

It is clear that $(\bar{c}b)_{V-A}$ and $(\bar{s}b)_{V-A} $ in Eq. (\ref{1na}) belong to the singlet and triplet representations of $SU(3)$, respectively. Thus in the factorization ansatz, only possible decays of $\Lambda_b$ belonging to $\bar{3}$ representation of $SU(3)$ are $\Lambda_b \to \Lambda_c^{+}D^{-}_{s},\;\Lambda_c^{+}\pi^{-}$ for the first term in Eq. (\ref{1na}).  For the second term, since $\bar{3}\times 3 = 8 + 1$, the possible decay mode of $\Lambda_b$ is $\Lambda_b \to \Lambda J/\psi$, as $\Lambda$ belongs to octet representation of $SU(3)$. 

Hence for the decay $\Lambda_b\left(\frac{1}{2}^{+}\right) \to \mathcal{B}_c^{*}\left(\frac{3}{2}^{+}\right)+P_{\pi}$ and $\Lambda_b\left(\frac{1}{2}^{+}\right) \to \mathcal{B}^{*}\left(\frac{3}{2}^{+}\right)+P_{D}$, where $\mathcal{B}_c^{*}\left(\frac{3}{2}^{+}\right)$ and $\mathcal{B}^{*}\left(\frac{3}{2}^{+}\right)$ belong to sextet and decuplet representations of $SU(3)$, respectively, the above framework is not applicable.

For these decays the dominant contribution comes from $W$ exchange.


$W$ exchange in the non-relativistic limit is encoded in the effective
Hamiltonian \cite{key-1,key-2,key-3}:

\begin{equation}
\mathcal{H}_{W}^{PC}=\frac{G_{F}}{\sqrt{2}}V_{ud}V_{cb}\underset{i\neq j}{\sum}\alpha_{i}^{-}\gamma_{j}^{+}(1-{\bf \sigma}_{i}\cdot{\bf \sigma}_{j})\delta^{3}(r),\label{eq:2}
\end{equation}
where 

\begin{align*}
\alpha_{i}^{-}\left|u\right\rangle  & =\left|d\right\rangle ,\quad i=1,2\text{ to be summed over \ensuremath{i}}\\
\gamma_{j}^{+}\left|b\right\rangle  & =\left|c\right\rangle ,\ j=3.
\end{align*}

Note that $\varLambda_{b},\varLambda_{c}^{+},\Xi_{c}^{0}$ belong
to anti-symmetric representation $\bar{3}$ of $SU(3)$, with spin
wave function 

\begin{equation}
\chi_{\text{MA}}^{\frac{1}{2},\frac{1}{2}}=\frac{1}{\sqrt{2}}\left|\left(\uparrow\downarrow-\downarrow\uparrow\right)\uparrow\right\rangle ,\label{eq:3}
\end{equation}
where as $\Sigma_{c}^{0},\Sigma_{b}^{+}$ belong to symmetric representation
of $6$ of $SU(3)$ with spin wave function

\begin{equation}
\chi_{\text{MS}}^{\frac{1}{2},\frac{1}{2}}=\frac{1}{\sqrt{6}}\left|-\left(\uparrow\downarrow+\downarrow\uparrow\right)\uparrow+2\uparrow\uparrow\downarrow\right\rangle .\label{eq:4}
\end{equation}
From the above equation, it follows 

\begin{align}
\left[\alpha_{1}^{-}\gamma_{3}^{+}\left(1-{\bf \sigma}_{1}\cdot{\bf \sigma}_{3}\right)+\alpha_{2}^{-}\gamma_{3}^{+}\left(1-{\bf \sigma}_{2}\cdot{\bf \sigma}_{3}\right)\right]\left|\varLambda_{b}\right\rangle  & =\sqrt{6}\left|\Sigma_{c}^{0}\right\rangle ,\label{eq:5}\\
\left[\alpha_{2}^{-}\gamma_{3}^{+}\left(1-{\bf \sigma}_{2}\cdot{\bf \sigma}_{3}\right)+\alpha_{1}^{-}\gamma_{3}^{+}\left(1-{\bf \sigma}_{1}\cdot{\bf \sigma}_{3}\right)\right]\left|\Sigma_{b}^{+}\right\rangle  & =\sqrt{6}\left|\varLambda_{c}^{+}\right\rangle ,\label{eq:6}\\
\left[\alpha_{2}^{-}\gamma_{3}^{+}\left(1-{\bf \sigma}_{2}\cdot{\bf \sigma}_{3}\right)+\alpha_{1}^{-}\gamma_{3}^{+}\left(1-{\bf \sigma}_{1}\cdot{\bf \sigma}_{3}\right)\right]\left|\Sigma_{b}^{+}\right\rangle  & =3\sqrt{2}\left|\Sigma_{c}^{+}\right\rangle .\label{eq:7}
\end{align}
The following comment is in order with respect to Eq. (\ref{eq:6})
and Eq. (\ref{eq:7}): one notes
\[
6\times8=\bar{3}+6+15+24.
\]
$W$ exchange is relevant when one considers baryon-pole contributions
(Born terms) involving the matrix elements of the form $\left\langle \mathcal{B}_{c}\left|\mathcal{H}_{W}^{PC}\right|\mathcal{B}_{b}\right\rangle ,$
which can be evaluated in the non-relativistic quark model (NQM) by
using Eq. (\ref{eq:2}). The form of Hamiltonian Eq. (\ref{eq:2})
lead to non-zero matrix elements \cite{key-1}:

\begin{equation}
\left\langle \Sigma_{c}^{0}\left|\mathcal{H}_{W}^{PC}\right|\Lambda_{b}\right\rangle =\left(\frac{G_{F}}{\sqrt{2}}V_{ud}V_{cb}\right)\sqrt{6}d^{\prime},\label{eq:8}
\end{equation}

\begin{equation}
\left\langle \Lambda_{c}^{+}\left|\mathcal{H}_{W}^{PC}\right|\Sigma_{b}^{+}\right\rangle =\left(\frac{G_{F}}{\sqrt{2}}V_{ud}V_{cb}\right)\sqrt{6}d^{\prime},\label{eq:9}
\end{equation}

\begin{equation}
\left\langle \Sigma_{c}^{-}\left|\mathcal{H}_{W}^{PC}\right|\Sigma_{b}^{+}\right\rangle =\left(\frac{G_{F}}{\sqrt{2}}V_{ud}V_{cb}\right)3\sqrt{2}d^{\prime}.\label{eq:10}
\end{equation}
where \cite{key-1,key-4}

\begin{equation}
d^{\prime}=\frac{3(m_{\Delta}-m_{N})}{8\pi\alpha_{s}}m_{bd}^{2}\approx10^{-2}\text{GeV}^{3}.\label{eq:11}
\end{equation}
On using the numerical values,

\begin{align*}
\alpha_{s}=0.32, & m_{b}=4.66\text{ \ensuremath{\text{GeV}}},\\
m_{d}\approx m_{u}=0.336\text{GeV}, & m_{s}=0.510\text{ \ensuremath{\text{GeV}},}\\
m_{c}=1.43\text{ \ensuremath{\text{GeV}}}, & m_{bd}=\frac{m_{b}m_{d}}{m_{b}+m_{d}}\approx0.313\text{ \ensuremath{\text{GeV}}},\\
m_{cu}=0.273\text{ \ensuremath{\text{GeV}}}, & m_{bs}\approx0.460\text{ \text{GeV}}.
\end{align*}

Hence, using $d^{\prime}$ given in Eq. (\ref{eq:11}) and $V_{cb}\approx0.040$
\begin{equation}
\frac{G_{F}}{\sqrt{2}}\left(V_{ud}V_{cb}\right)\sqrt{6}d^{\prime}\approx8.08\times10^{-9}\text{ \ensuremath{\text{GeV}}}.\label{eq:12}
\end{equation}

\section{Hadronic decays of \textmd{$\mathcal{B}_{b}^{\prime}\left(\frac{1}{2}^{+}\right)\to\mathcal{B}^{*}\left(\frac{3}{2}^{+}\right)+P_{D}$}}

For the decay of above type, the decay rate is given by \cite{key-1}
\begin{equation}
\Gamma=\frac{1}{6\pi}\frac{m^{\prime}}{m^{*2}}\frac{\left|{\bf p}\right|^{3}}{f_{D}^{2}}(p_{0}+m^{*})\left|C\right|^{2},\label{eq:13}
\end{equation}
where $C$ is the parity conserving (p-wave amplitude). Thus for the
decays $\Lambda_{b}\to\Delta^{0}D^{0},\Delta^{-}D^{+},\Sigma^{*-}D_{s}^{+},$
Using $f_{D}\approx0.207\text{ GeV}$, $f_{D_{s}^{+}}\approx0.257\text{ GeV}$,
from Eq. (\ref{eq:13})
\begin{align}
\Gamma(\Lambda_{b}\to\Delta^{0}D^{0}) & \approx3.99\times10^{2}\left|C\right|^{2}\text{ GeV},\label{eq:14}\\
\Gamma(\Lambda_{b}\to\Delta^{-}D^{+}) & \approx3.99\times10^{2}\left|C\right|^{2}\text{ GeV},\label{eq:15}\\
\Gamma(\Lambda_{b}\to\Sigma^{*-}D_{s}^{+}) & \approx1.31\times10^{2}\left|C\right|^{2}\text{ GeV.}\label{eq:16}
\end{align}
In order to determine the amplitude $C$, one notes
\begin{align}
F_{i}T_{ijk}S^{jk} & \to\left[F_{i}T_{i22}\right]S^{22},\nonumber \\
 & =\left[F_{1}T_{122}+F_{2}T_{222}+F_{3}T_{322}\right]S^{22},\nonumber \\
 & =\left[\Delta^{0}D^{0}+\sqrt{3}\Delta^{-}D^{+}+\Sigma^{*-}D_{s}^{+}\right]2\Sigma_{c}^{0},\label{eq:17}
\end{align}
where $F_{i}=\left(D^{0},D^{+},D_{s}^{+}\right)$

Thus for the decays $\Lambda_{b}\to\Delta^{0}D^{0},\Delta^{-}D^{+},\Sigma^{*-}D_{s}^{+}$
\cite{key-1}
\begin{equation}
C=\left(1,\sqrt{3},1\right)2g^{*}\frac{\left\langle \Sigma_{c}^{0}\left|\mathcal{H}_{W}^{PC}\right|\Lambda_{b}\right\rangle }{m_{\Lambda_{b}}-m_{\Sigma_{c}^{0}}}.\label{eq:18}
\end{equation}
Hence from Eq. (\ref{eq:8}) and Eq. (\ref{eq:12})
\begin{equation}
C=\left(1,\sqrt{3},1\right)\left(4.33\right)\times10^{-9},\label{eq:19a}
\end{equation}
where we have used \cite{key-1}
\begin{equation}
2g^{*}=1.70.\label{eq:20}
\end{equation}
From Eq. (\ref{eq:14},\ref{eq:15},\ref{eq:16}) and Eq. (\ref{eq:19a}),
we get
\begin{align}
\text{Br}(\Lambda_{b}\to\Delta^{0}D^{0}) & \approx1.7\times10^{-2},\label{eq:21a}\\
\text{Br}(\Lambda_{b}\to\Delta^{-}D^{+}) & \approx5.0\times10^{-2},\label{eq:22}\\
\text{Br}(\Lambda_{b}\to\Sigma^{*-}D_{s}^{+}) & \approx5.4\times10^{-3}.\label{eq:23}
\end{align}
To take into $SU(3)$ breaking, Eq. (\ref{eq:23}) is multiplied by
a factor $\left(\frac{m_{bs}}{m_{bd}}\right)^{2}\approx2.16$. Thus,
\begin{equation}
\text{Br}(\Lambda_{b}\to\Sigma^{*-}D_{s}^{+})\approx1.2\times10^{-2}.\label{eq:24}
\end{equation}

\section{Hadronic decays of \textmd{$\mathcal{B}_{b}^{\prime}\left(\frac{1}{2}^{+}\right)\to\mathcal{B}_c^{*}\left(\frac{3}{2}^{+}\right)+P$}}

For the decays $\Lambda_{b}\to\Sigma_{c}^{*+}\pi^{-},\Sigma_{c}^{*0}\pi^{0},\Xi_{c}^{*0}K^{0}$,
from Eq. (\ref{eq:13}), using $f_{\pi}=0.130$ GeV, $f_{K}=0.161$GeV,
we have 
\begin{align}
\Gamma\left(\Lambda_{b}\to\Sigma_{c}^{*+}\pi^{-}\right)=\Gamma\left(\Lambda_{b}\to\Sigma_{c}^{*0}\pi^{0}\right) & \approx2.14\times10^{2}\left|C\right|^{2}\text{ \ensuremath{GeV}},\label{eq:25}\\
\Gamma(\Lambda_{b}\to\Xi_{c}^{*0}K^{0}) & \approx1.00\times10^{2}\left|C\right|^{2}\text{ \ensuremath{GeV}}.\label{eq:26}
\end{align}
In order to determine the amplitude $C$, note that
\begin{align}
S_{ij}^{*}P_{k}^{j}S^{ik} & \to\left[S_{2j}^{*}P_{2}^{j}\right]S^{22}\nonumber \\
 & =\left[\Sigma_{c}^{*+}\pi^{-}-\Sigma_{c}^{*0}\pi^{0}+\Xi_{c}^{*0}K^{0}\right]\sqrt{2}\Sigma_{c}^{0}.\label{eq:27}
\end{align}
Thus the amplitude $C$ for the decays $\Lambda_{b}\to\Sigma_{c}^{*+}\pi^{-},\Sigma_{c}^{*0}\pi^{0},\Xi_{c}^{*0}K^{0}$
is given by \cite{key-1}
\begin{equation}
C=\left(1,-1,1\right)\sqrt{2}g_{c}^{*}\left(2.55\times10^{-9}\right).\label{eq:28}
\end{equation}
Now the decay rate for the decay $S^{*}\to A\pi$ is given by
\begin{equation}
\Gamma\left(S^{*}\to A\pi\right)=\frac{1}{\sqrt{2\pi}}\frac{1}{m_{S}^{*}}\frac{\left(\sqrt{2}g_{c}^{*}\right)^{2}}{f_{\pi}^{2}}(p_{0}+m_{A})\left|{\bf p}\right|^{3}.\label{eq:29}
\end{equation}
The $SU(3)$ symmetry gives: $\sqrt{2}g_{c}^{*}\left[\Sigma_{c}^{*++}\pi^{+}-\Sigma_{c}^{*+}\pi^{0}\right]\Lambda_{c}^{+}.$

From Eq. (\ref{eq:29}), we get
\begin{equation}
\Gamma\left(\Sigma_{c}^{*++}\to\Lambda_{c}^{+}\pi^{+}\right)=(\sqrt{2}g_{c}^{*})^{2}(1.673),\tag{30}\label{eq:30}
\end{equation}
From the experimental value for this decay
\begin{equation}
\sqrt{2}g_{c}^{*}\approx0.94.\label{eq:31}
\end{equation}
Thus from Eq. (\ref{eq:28}) and Eq. (\ref{eq:31}),
\begin{equation}
C=\left(1,-1,1\right)\left(2.40\times10^{-9}\right),\label{eq:32}
\end{equation}
Hence from Eq. (\ref{eq:25}, \ref{eq:26}) and Eq. (\ref{eq:32})
\begin{align}
\text{Br}(\Lambda_{b}\to\Sigma_{c}^{*+}\pi^{-})=\text{Br}\left(\Lambda_{b}\to\Sigma_{c}^{*0}\pi^{0}\right) & \approx2.74\times10^{-3},\label{eq:33}\\
\text{Br}(\Lambda_{b}\to\Xi_{c}^{*0}K^{0}) & \approx1.28\times10^{-3},\label{eq:34}\\
 & \to\left(1.28\times10^{-3}\right)\left(\frac{m_{bs}}{m_{bd}}\right)^{2},\nonumber \\
 & \approx2.76\times10^{-3}.\label{eq:35}
\end{align}

\section{Hadronic Weak Decays \cite{key-5}}

For the hadronic weak decay $\mathcal{B}^{\prime}\left(p^{\prime}\right)\to\mathcal{B}\left(p\right)+\pi\left(k\right)$,
with $p^{'}=p+k$, the transition matrix can be written as
\begin{equation}
T=\frac{1}{\left(2\pi\right)^{\frac{3}{2}}}\sqrt{\frac{mm^{\prime}}{2k_{0}p_{0}p_{0}^{\prime}}}\bar{u}\left(p\right)\left(A-B\gamma_{5}\right)u\left(p^{\prime}\right)\label{eq:36-1}
\end{equation}
where $A$ and $B$ are the parity violating (s-wave) and parity conserving
(p-wave) amplitudes, respectively. The Decay rate is given by
\begin{equation}
\Gamma=\frac{k}{4\pi m^{\prime}}\left[\left(p_{0}+m\right)\left|A\right|^{2}+\left(p_{0}-m\right)\left|B\right|^{2}\right],\label{eq:37-1}
\end{equation}
and the asymmetry parameter 
\begin{equation}
\alpha=\frac{2kA^{*}B}{\left[\left(p_{0}+m\right)\left|A\right|^{2}+\left(p_{0}-m\right)\left|B\right|^{2}\right]}.\label{eq:38-1}
\end{equation}

\subsection{Factorization}

For the decay $\Lambda_{b}\left(p^{\prime}\right)\to\Lambda_{c}^{+}\left(p\right)\pi^{-}\left(k\right)$,
the factorization gives
\begin{equation}
\left(\frac{G_{F}}{\sqrt{2}}V_{ud}V_{cb}\right)a_{1}\left(-f_{\pi}\right)k^{\mu}\bar{u}\left(p\right)\gamma_{\mu}\left(g_{V}-g_{A}\gamma_{5}\right)u\left(p^{\prime}\right).\label{eq:40-1}
\end{equation}
Thus
\begin{align}
A_{\text{fact}} & =-\left(\frac{G_{F}}{\sqrt{2}}V_{ud}V_{cb}\right)a_{1}f_{\pi}\left(m_{\Lambda_{b}}-m_{\Lambda_{c}}\right)g_{V},\label{eq:41-1}\\
B_{\text{fact}} & =\left(\frac{G_{F}}{\sqrt{2}}V_{ud}V_{cb}\right)a_{1}f_{\pi}\left(m_{\Lambda_{b}}+m_{\Lambda_{c}}\right)g_{A}.\label{eq:42-1}
\end{align}
The quark model gives $g_{V}=1,\quad g_{A}=0,$ implying $B_{\text{fact}}=0.$
Now
\[
a_{1}=C_{1}+\frac{1}{3}C_{2},\quad C_{1}=1.009,\quad C_{2}=-0.257,
\]
giving $a_{1}=0.92.$ The $g_{V}=1$, is in the limit $m\to0$, i.e.,
$g_{V}\left(0\right)=1$. To get $g_{V}$ for finite masses, the following
prescription is used
\begin{equation}
g_{V}=\left(1+\frac{m_{d}}{m_{b}}\right)\left(1-\frac{m_{c}}{m_{b}}\right)g_{V}\left(0\right)=0.739\label{eq:43-1}
\end{equation}
and 
\begin{equation}
a_{1}g_{V}=0.680\label{eq:44-1}
\end{equation}
 by using $m_{d}=0.336$ GeV, $m_{c}=1.45$ GeV, $m_{b}=4.66$ GeV.
Hence from Eqs. (\ref{eq:41-1}) and (\ref{eq:44-1}), we have
\begin{equation}
A=A_{\text{fact}}=-9.47\times10^{-8},\label{eq:45-1}
\end{equation}
and from Eqs. (\ref{eq:37-1}), using $k=2.342$ GeV, $p_{0}+m_{\Lambda_{c}}=5.558$
GeV, we have
\begin{equation}
\text{Br}\left(\Lambda_{b}\to\Lambda_{c}^{+}\pi^{-}\right)=4.0\times10^{-3},\label{eq:46-1}
\end{equation}
to be compared with the experimental value
\begin{equation}
\text{Br}\left(\Lambda_{b}\to\Lambda_{c}^{+}\pi^{-}\right)=\left(4.9\pm0.4\right)\times10^{-3}.\label{eq:47}
\end{equation}

For the decay $\Lambda_{b}\to\Lambda_{c}^{+}D_{s}^{-}$, we can write
\begin{equation}
A=A_{\text{fact}}=-\left(\frac{G_{F}}{\sqrt{2}}V_{ud}V_{cb}\right)a_{1}g_{V}f_{D_{s}}\left(m_{\Lambda_{b}}-m_{\Lambda_{c}}\right)=1.92\times10^{-7}\label{eq:48}
\end{equation}
on using $f_{D_{s}}=0.257$ GeV. 

Hence, from Eq. (\ref{eq:37-1}), using $k=1.833$ GeV, $p_{0}+m_{\Lambda_{c}}=5.216$
GeV, we have
\begin{equation}
\text{Br}\left(\Lambda_{b}\to\Lambda_{c}^{+}D_{s}^{-}\right)=1.11\times10^{-2}=1.11\%\label{eq:49}
\end{equation}
in remarkable agreement with the experimental value
\[
\text{Br}\left(\Lambda_{b}\to\Lambda_{c}^{+}D_{s}^{-}\right)=\left(1.10\pm0.10\right)\%.
\]

Although, $B_{\text{fact}}=0$ for $\Lambda_{b}\to\Lambda_{c}^{+}\pi^{-}$
- but there is a dominant contribution for the amplitude $B$ from
$W$ exchange, i.e., from the baryon pole:
\begin{equation}
B=B\text{(pole)}=g_{\Lambda_{b}\pi^{-}\Sigma_{b}^{+}}\frac{1}{m_{\Sigma_{b}^{+}}-m_{\Lambda_{c}^{+}}}\left\langle \Lambda_{c}^{+}\left|\mathcal{H}_{W}^{PC}\right|\Sigma_{b}^{+}\right\rangle .\label{eq:36}
\end{equation}
The Goldberger-Treiman (GT) relation gives
\begin{align}
g_{\Lambda_{b}\pi^{-}\Sigma_{b}^{+}} & =\frac{m_{\Lambda_{b}}+m_{\Sigma_{b}^{+}}}{f_{\pi}}g_{A},\label{eq:38}
\end{align}
The quark model gives
\begin{equation}
g_{A}=\sqrt{\frac{2}{3}}\label{eq:40}
\end{equation}
and from Eqs. (\ref{eq:9}, \ref{eq:12}) 
\[
\left\langle \Lambda_{c}^{+}\left|\mathcal{H}_{W}^{PC}\right|\Sigma_{b}^{+}\right\rangle =\left(8.08\times10^{-9}\right)\text{\text{ GeV}},
\]
hence
\begin{align}
B\text{(pole)} & =\left(\frac{m_{\Lambda_{b}}+m_{\Sigma_{b}^{+}}}{m_{\Sigma_{b}^{+}}-m_{\Lambda_{c}^{+}}}\right)\frac{g_{A}}{f_{\pi}}\left(8.08\times10^{-9}\text{ GeV}\right),\nonumber \\
 & =1.67\times10^{-7}\text{GeV}.\label{eq:41}
\end{align}
Again, the quark model value $g_{A}=\sqrt{\frac{2}{3}}$ is in the
limit $m\to0.$ To get $g_{A}$ for the finite mass, the following
parameterization is used:
\begin{equation}
g_{A}=\left(1-\frac{m_{c}}{m_{b}}\right)g_{A}\left(0\right)=0.689\sqrt{\frac{2}{3}.}\label{eq:42-2}
\end{equation}
Thus 
\[
B\left(\text{pole}\right)=\left(1.67\times10^{-7}\right)0.689\text{ GeV}=1.10\times10^{-7}\text{GeV.}
\]
Hence, from Eq. (\ref{eq:37-1}), using $p_{0}+m_{\Lambda_{c}}=0.99\text{ GeV},$we
have
\begin{equation}
\text{Br}\left(\Lambda_{b}\to\Lambda_{c}^{+}\pi^{-}\right)=4.9\times10^{-3},\label{eq:43-2}
\end{equation}
 in remarkable agreement with the corresponding experimental value
\begin{equation}
\text{Br}\left(\Lambda_{b}\to\Lambda_{c}^{+}\pi^{-}\right)=\left(4.9\pm0.4\right)\times10^{-3}.\label{eq:44-2}
\end{equation}
From Eq. (\ref{eq:38-1}), the asymmetry parameter $\alpha=-0.77$. 

The non-leptonic decay $\Lambda_{b}\to\Sigma_{c}^{+}\pi^{-}$ is of
considerable interest. Factorization does not contribute to this decay
and only baryon pole contributes:
\begin{equation}
g_{\Lambda_{b}\Sigma_{b}^{+}\pi^{-}}=\frac{1}{m_{\Lambda_{b}}-m_{\Sigma_{c}^{+}}}\left\langle \Sigma_{c}^{+}\left|\mathcal{H}_{W}^{PC}\right|\Sigma_{b}^{+}\right\rangle .\label{eq:45-2}
\end{equation}
GT relation gives
\begin{equation}
g_{\Lambda_{b}\Sigma_{b}^{+}\pi^{-}}=\frac{m_{\Lambda_{b}}+m_{\Sigma_{b}^{+}}}{f_{\pi}}g_{A},\quad g_{A}=\sqrt{\frac{2}{3}}.\label{eq:46-2}
\end{equation}
Thus

\begin{align}
B=B\text{(pole)} & =\left(\frac{m_{\Lambda_{b}}+m_{\Sigma_{b}^{+}}}{m_{\Lambda_{b}}-m_{\Sigma_{c}^{+}}}\right)\frac{g_{A}}{f_{\pi}}\left\langle \Sigma_{c}^{+}\left|\mathcal{H}_{W}^{PC}\right|\Sigma_{b}^{+}\right\rangle .\label{eq:47-1}
\end{align}
From Eqs. (\ref{eq:10}) and (\ref{eq:12}):
\begin{equation}
\left\langle \Sigma_{c}^{+}\left|\mathcal{H}_{W}^{PC}\right|\Sigma_{b}^{+}\right\rangle =\sqrt{3}\left(8.08\times10^{-9}\right)\text{GeV}.\label{eq:48-1}
\end{equation}
By using $m_{\Sigma_{c}^{+}}=2.45$ GeV, from Eq. (\ref{eq:47-1}),
we have
\[
B=B\left(\text{pole}\right)=3.12\times10^{-7},\quad p_{0}-m_{\Sigma_{c}^{+}}=0.89\text{ GeV}.
\]
Hence, from Eq. (\ref{eq:37-1}), one gets
\begin{equation}
\text{Br}\left(\Lambda_{b}\to\Sigma_{c}^{+}\pi^{-}\right)=2.9\times10^{-3}.\label{eq:49-1}
\end{equation}

To conclude: in this work, the formalism developed in \cite{key-1}is
extended to $\Lambda_{b}$ decays
\begin{align*}
\Lambda_{b} & \to\Delta^{0}D^{0},\Delta^{-}D^{+},\Sigma_{c}^{*-}D_{s}^{+}\\
\Lambda_{b} & \to\Sigma_{c}^{*+}\pi^{-},\Sigma^{*0}\pi^{0},\Xi_{c}^{*0}K^{0}.
\end{align*}
The experimental data is not available to check our results. The most
interesting decay in this category is

\[
\Lambda_{b}\to\Delta^{0}D^{0}\to p\pi^{-}D^{0},
\]
The experimental value for the branching ratio for the decay $\Lambda_{b}\to p\pi^{-}D^{0}$
\cite{key-6} is
\begin{equation}
\text{Br}\left(\Lambda_{b}\to p\pi^{-}D^{0}\right)=\left(6.4\pm0.7\right)\times10^{-4},\label{eq:45}
\end{equation}
where as, our result of the branching ratio is
\[
\text{Br}\left(\Lambda_{b}\to\Delta^{0}D^{0}\right)\approx1.7\times10^{-2}.
\]
Thus
\begin{equation}
\frac{\text{Br}\left(\Lambda_{b}\to p\pi^{-}D^{0}\right)}{\text{Br}\left(\Lambda_{b}\to\Delta^{0}D^{0}\right)}\approx3.9\times10^{-2}.\label{eq:46}
\end{equation}
One notes
\[
\Gamma(\Delta^{0}\to p\pi^{-})\approx39\text{ MeV}.
\]

For the decay
\[
\Lambda_{b}\to\Lambda_{c}^{+}D_{s}^{-}
\]
$B_{\text{fact}}=0$ and $B\left(\text{pole}\right)=0$. Thus for
this decay only factorization contributes. The branching ratio
\[
\text{Br}\left(\Lambda_{b}\to\Lambda_{c}^{+}D_{s}^{-}\right)=1.11\%
\]
is in agreement with the experimental value, where as, for the decay
$\Lambda_{b}\to\Lambda_{c}^{+}\pi^{-}$, the factorization gives
\[
\text{Br}\left(\Lambda_{b}\to\Lambda_{c}^{+}\pi^{-}\right)=4.0\times10^{-3},\text{ and }\alpha=0,
\]
which is not in agreement with the experimental value. However, after
including the baryon pole contribution to the amplitude $B$, gives
the branching ratio
\[
\text{Br}\left(\Lambda_{b}\to\Lambda_{c}^{+}\pi^{-}\right)=4.9\times10^{-3}
\]
and the asymmetry parameter to be
\[
\alpha=-0.77.
\]
This result of the branching ratio is in agreement with the experimental
value.

The branching ratio for the decay $\Lambda_{b}\to\Sigma_{c}^{+}\pi^{-}\to\Lambda_{c}^{+}\pi^{0}\pi^{-}$
is
\[
\text{Br}\left(\Lambda_{b}\to\Sigma_{c}^{+}\pi^{-}\right)\approx2.9\times10^{-3}
\]
and may be of an interest for the experimentalists. 

The decays for which the experimental data is available, our results
for the branching ratios are in agreement with them. The experimental
verification of our results for the asymmetry parameter $\alpha$
and the branching ratio for the decay $\Lambda_{b}\to\Sigma_{c}^{+}\pi^{-}$
, would give an important boost to the formalism developed in \cite{key-1}
for the non-leptonic decays of $\Lambda_{b}$.

\newpage{}

\end{document}